\documentclass[sigconf]{acmart}
\usepackage{multirow}
\usepackage{enumitem}
\usepackage{booktabs}
\usepackage{makecell}

\AtBeginDocument{%
  \providecommand\BibTeX{{%
    \normalfont B\kern-0.5em{\scshape i\kern-0.25em b}\kern-0.8em\TeX}}}

\setcopyright{acmlicensed}
\copyrightyear{2018}
\acmYear{2018}
\acmDOI{XXXXXXX.XXXXXXX}

\acmConference[Conference acronym 'XX]{Make sure to enter the correct
  conference title from your rights confirmation emai}{June 03--05,
  2018}{Woodstock, NY}
\acmISBN{978-1-4503-XXXX-X/18/06}

\begin{document}

\title{Towards Unified Modeling for Positive and Negative Preferences in Sign-Aware Recommendation}

\author{Yuting Liu}
\affiliation{%
  \institution{Software College, Northeastern University}
  \country{China}
}
\email{yutingliu@stumail.neu.edu.cn}

\author{Yizhou Dang}
\affiliation{%
  \institution{Software College, Northeastern University}
  \country{China}
}
\email{dangyz@stumail.neu.edu.cn}

\author{Yuliang Liang}
\affiliation{%
  \institution{Software College, Northeastern University}
  \country{China}
}
\email{liangyuliang@stumail.neu.edu.cn}

\author{Qiang Liu}
\affiliation{%
  \institution{Center for Research on Intelligent Perception and Computing, Institute of Automation, Chinese Academy of Sciences}
  \country{China}
}
\email{qiang.liu@nlpr.ia.ac.cn}

\author{Guibing Guo}
\authornote{Corresponding author.}
\affiliation{%
  \institution{Software College, Northeastern University}
  \country{China}
}
\email{guogb@swc.neu.edu.cn}

\author{Jianzhe	Zhao}
\affiliation{%
  \institution{Software College, Northeastern University}
  \country{China}
}
\email{zhaojz@swc.neu.edu.cn}

\author{Xingwei Wang}
\affiliation{%
  \institution{Northeastern University}
  \country{China}
}
\email{wangxw@swc.neu.edu.cn}

\renewcommand{\shortauthors}{Trovato and Tobin, et al.}

\begin{abstract}
 Recently, sign-aware graph recommendation has drawn much attention as it will learn users' negative preferences besides positive ones from both positive and negative interactions (i.e., links in a graph) with items.  
 To accommodate the different semantics of negative and positive links, existing works utilize two independent encoders to model users' positive and negative preferences, respectively. However, these approaches cannot learn the negative preferences from high-order heterogeneous interactions between users and items formed by multiple links with different signs, resulting in inaccurate and incomplete negative user preferences.
 To cope with these intractable issues, we propose a novel \textbf{L}ight \textbf{S}igned \textbf{G}raph Convolution Network specifically for \textbf{Rec}ommendation (\textbf{LSGRec}), which adopts a unified modeling approach to simultaneously model high-order users' positive and negative preferences on a signed user-item interaction graph.
 Specifically, for the negative preferences within high-order heterogeneous interactions, first-order negative preferences are captured by the negative links, while high-order negative preferences are propagated along positive edges.
 Then, recommendation results are generated based on positive preferences and optimized with negative ones. Finally, we train representations of users and items through different auxiliary tasks.
 Extensive experiments on three real-world datasets demonstrate that our method outperforms existing baselines regarding performance and computational efficiency. Our code is available at \url{https://anonymous.4open.science/r/LSGRec-BB95}.
\end{abstract}





\maketitle

\section{Introduction}
Due to its powerful ability to model user-item interactions in a bipartite graph, graph-based collaborative filtering~\cite{ngcf,lightgcn,idsf,mmgcn,ultragcn,lattice} has become the mainstream method for recommendation systems. Graph-based approaches usually adopt a message-passing and neighborhood aggregation mechanism in user-item bipartite graphs to capture high-order collaborative signals that model users' preferences and learn effective user and item representations. 
Most existing works in graph-based recommendations~\cite{Gori2007ItemRankAR,Defferrard2016ConvolutionalNN} treat all user-item interactions as positive, ignoring actual negative interactions between users and items that also reflect users’ personalized preferences in real-world recommendation systems (e.g., a user gives a negative review after purchasing an item)~\cite{negpos,Wang2023LearningFN}. Without considering this type of interaction, these methods are likely to mistakenly treat items that users dislike as they like, resulting in inaccurate user preferences.

\begin{figure*}[t]
    \centering
    \includegraphics[width=\linewidth]{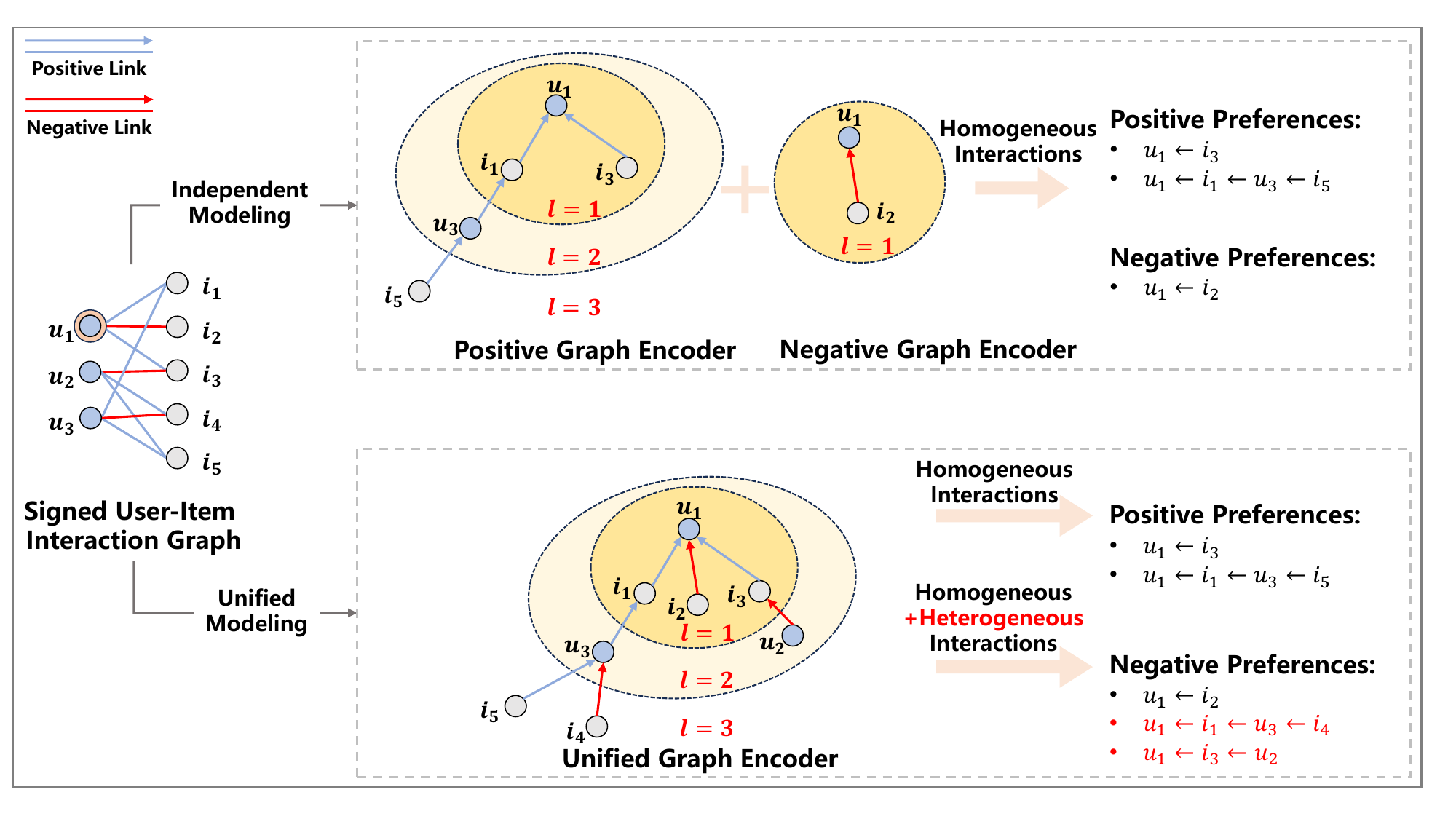}
    \caption{An illustration of the user-item signed interaction graph and its high-order connectivity through unified modeling and independent modeling, respectively. The node $u_1$ is the target user that needs recommendations. The rightmost side shows the paths of positive and negative preferences in different modeling approaches.}
    \label{fig: example}
\end{figure*}

Recently, a few studies have paid attention to explicitly modeling users' negative preferences besides traditional positive ones in the graph-based recommendation~\cite{negpos,pane-gnn,siren,Olivier2019Revisiting,Huang2023DualLightGCN}. Since the negative links inherently have different semantics and principles compared to positive links~\cite{sgcn,sigat,Fan2019GraphNN,CAI2022102917,ghcf}, they split interactions into positive and negative and model users' positive and negative preferences with corresponding independent encoders, respectively. Although effective, these methods cause some problems. Intuitively, they may disrupt high-order collaborative signals in the signed graph. As shown in Fig~\ref{fig: example}, $u_1$ and $u_2$ interact with the same item $i_3$, but $u_1$ likes $i_3$, and $u_2$ does not, we can assume that the preference of $u_1$ is negative correlative with the preference of $u_2$. This collaborative signal that the preferences of $u_1$ and $u_2$ are dissimilar can be captured by unified modeling but cannot be captured by existing independent modeling approaches.
Furthermore, independent modeling for users and items on two separate graphs ignores the preferences within high-order heterogeneous interactions (i.e., the links between users and items formed by multiple edges with different signs). Taking $u_1$ in Fig~\ref{fig: example} as an example, we can observe that there are two additional paths for high-order negative preferences when adopting unified modeling for users and items (i.e., $u_1\leftarrow i_1\leftarrow u_3\leftarrow i_4$ and $u_1\leftarrow i_3\leftarrow u_2$ in red font) compared to adopting independent modeling. Users' preferences will be incomplete and inaccurate without considering these paths, which prompts us to propose a new unified modeling approach.

In this paper, we propose \textbf{LSGRec}, a \textbf{L}ight \textbf{S}igned \textbf{G}raph Convolution Network for \textbf{Rec}ommendation to achieve these, including an effective unified modeling method to calculate user/item representations, a negative preference filter to generate corrected recommendations and multiple training objectives to optimize the parameters.
Specifically, we devise a unified modeling approach to model high-order users' positive and negative preferences simultaneously on a signed user-item interaction graph. For the negative preferences within high-order heterogeneous interactions, first-order negative preferences are captured by the negative links, while high-order negative preferences are propagated along positive edges based on homophily. We combine the representations learned at different propagation hops to obtain the final positive and negative embeddings for prediction. 

Then, we propose a negative preference filter to generate revised recommendations and multiple training objectives to optimize the embeddings of users and items, including sign-aware Bayesian personalized ranking, link prediction, and an orthogonality constraint to force positive and negative representations to share no common information. To sum up, the main contributions of our work are the following:

\begin{itemize}
    \item To make up for the negative preferences within high-order heterogeneous interactions, we devise a simple and elegant unified modeling approach to propagate and aggregate positive and negative preferences in the signed graph simultaneously.
    \item We propose LSGRec, a novel graph-based recommendation model that makes full use of positive and negative user-item interactions to provide more accurate recommendations, which consists of a unified modeling method, a negative preference filter, and a training strategy with multiple auxiliary tasks.
    \item We perform extensive experiments on three real-world datasets (Amazon-Beauty, Amazon-Book, Yelp2021) to demonstrate that our method outperforms the state-of-the-art graph-based recommendation methods. Our method outperforms the best baseline by up to $10.64\%$, $16.21\%$, and $15.60\%$ in terms of Precision$@10$, Recall$@10$, and NDCG$@10$, respectively.
\end{itemize}

\section{Methodology}

In this section, we introduce the notations used in the paper, explain the architecture of our LSGRec, and describe the optimization objective in detail.

\subsection{Preliminary}

\begin{figure}[htbp]
    \centering
    \includegraphics[width=\linewidth]{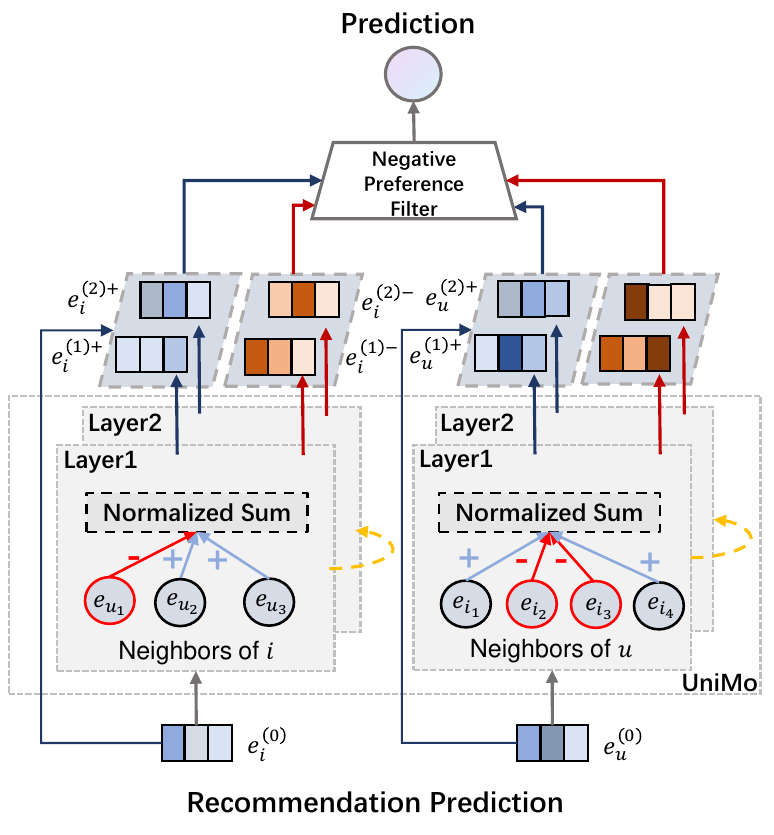}
    \vspace{-20pt}
    \caption{An illustration of our LSGRec framework, consisting of a unified modeling approach, a negative preference filter, and multiple auxiliary tasks. Specifically, positive and negative embeddings are obtained by the unified encoder from a whole graph and are fed into the negative preference filter to generate recommendations.}
    \label{fig:model}
\end{figure}

The basic input in recommendation methods is the historical user-item interactions with ratings, which is modeled as a weighted bipartite graph $\mathcal{G}=(\mathcal{U},\mathcal{I},\mathcal{E})$, where $\mathcal{U}$ and $\mathcal{I}$ are the set of $M$ users and $N$ items, respectively, and $\mathcal{E}$ is the set of weighted edges between $\mathcal{U}$ and $\mathcal{I}$. A weighted edge $(u,i,\omega_{ui})\in\mathcal{E}$ represents that a user $u\in\mathcal{U}$ gives a rating $\omega_{ui}$ to an item $i\in\mathcal{I}$. To simplify the setting, we assume $\mathcal{G}$ as a static network without repeated edges.

In our study, to fully utilize $\omega_{ui}$ (i.e., ratings), we set a threshold $\delta$ to split the original ratings into binary signals $\mathcal{E}^+$ and $\mathcal{E}^-$, where

\begin{equation}
    \begin{aligned}
        \mathcal{E}^+&=\{(u,i,1)|\mbox{sign}(\omega_{ui}-\delta)>0, (u,i,\omega_{ui})\in\mathcal{E}\}, \\
        \mathcal{E}^-&=\{(u,i,-1)|\mbox{sign}(\omega_{ui}-\delta)<0, (u,i,\omega_{ui})\in\mathcal{E}\}.
    \end{aligned}
    \label{eq:edge_split}
\end{equation}

Here, $\mathcal{E}^+\subset\mathcal{U}\times\mathcal{I}$ and $\mathcal{E}^-\subset\mathcal{U}\times\mathcal{I}$ denote the sets of positive and negative edges, respectively. The function $\mbox{sign}(\cdot)$ outputs the sign of the input, and $\delta$ can be determined according to the characteristics of a given dataset. Unlike existing works that divide the whole graph into two edge-disjoint sub-graphs according to the sign of edges, we process these two signals in the original whole graph with a unified encoder; that is, a user-item weighted bipartite graph becomes a user-item signed bipartite graph $\mathcal{G}=(\mathcal{U},\mathcal{I},\mathcal{E}^+,\mathcal{E}^-)$. Note that $\mathcal{E}^+\cap\mathcal{E}^-=\emptyset$ and $\mathcal{E}^+\cup\mathcal{E}^-=\mathcal{E}$, in other words, a user cannot have both positive and negative preferences for an item simultaneously. Hence, given a user-item interaction graph $\mathcal{G}=(\mathcal{U},\mathcal{I},\mathcal{E}^+,\mathcal{E}^-)$, our task is to generate top$K$ recommendations for each user.

\subsection{Model Overview}

The architecture of the LSGRec model is depicted in Fig~\ref{fig:model}, consisting of a unified modeling (UniMo) approach, a negative preference filter, and multiple auxiliary tasks to train parameters. The UniMo calculates the final embeddings $\mathbf{e}_{u}^{+}$, $\mathbf{e}_{i}^{+}$, $\mathbf{e}_{u}^{-}$, $\mathbf{e}_{i}^{-}$ corresponding to the positive and negative preferences of users and items from their initial embeddings $\mathbf{e}^{(0)}_{u}$ and $\mathbf{e}^{(0)}_{i}$ by aggregating high-order neighborhood information iteratively. The negative preference filter utilizes negative preferences to filter out items that users dislike, generating more accurate recommendations. The multi-task learning objectives aim to train user/item embeddings better by optimizing different loss functions.

\subsection{Unified Modeling for Positive and Negative Preferences}

The basic idea of GCN is to learn the representation of nodes by smoothing features over the graph~\cite{gcn1st,SimplifyingGCN}. To achieve this, it adopts message-passing to aggregate the features of neighbors as the new representation of a target node, which can be abstracted as:
\begin{eqnarray}
    \mathbf{e}^{(l+1)}_{u}=AGG(\mathbf{e}_{u}^{(l)},\{\mathbf{e}_{i}^{(l)}|\ i\in\mathcal{N}_u\}).
    \label{eq:agg}
\end{eqnarray}

The $AGG$ is an aggregation function -- the core of graph convolution -- that considers the $l$-th layer's representation of the target node and its neighbors. Although many graph-based recommendation approaches~\cite{gat,graphsage,graphtransformer} have specified the $AGG$ and perform well on unsigned user-item bipartite graphs, they are not adaptive for signed graphs and cannot model negative edges and capture negative neighborhood information. The primary challenges are that negative links have a different semantic meaning from positive links, their principles are inherently different, and they form complex relations with positive links.

Next, we build upon the message-passing to capture collaborative signals along the signed user-item bipartite graph, unifiedly modeling the positive and negative embeddings of users and items in the whole graph. We first illustrate the design of first-order propagation and then generalize it to high-order recursively. Note that there are some minor differences between the first-order and high-order propagation due to the particularity of negative edges.

\subsubsection{Direct Neighbors}

Intuitively, the positively interacted items provide direct evidence of a user's positive preference (i.e., what a user likes), and the negatively interacted items imply a user's negative preference (i.e., what a user dislikes). In the first order, we calculate positive and negative representations of each node by aggregating messages passed by the direct neighbors through positive and negative edges. We adopt a simple weighted sum aggregator and abandon the user of feature transformation and nonlinear activation with the $AGG$ function, which could be burdensome for collaborative filtering. The signed graph convolution operation (i.e., propagation rule) in the first layer in UniMo can be defined as:
\begin{equation}
    \begin{aligned}
        \mathbf{e}_u^{(1)+}=\frac{1}{\sqrt{|\mathcal{N}^+_{u}|}\sqrt{|\mathcal{N}^+_{i}|}} \sum_{i\in \mathcal{N}^+_{u}} \mathbf{e}_i^{(0)}, \\
        \mathbf{e}_u^{(1)-}=\frac{1}{\sqrt{|\mathcal{N}^-_{u}|}\sqrt{|\mathcal{N}^-_{i}|}} \sum_{i\in \mathcal{N}^-_{u}} \mathbf{e}_i^{(0)},
        \label{eq:u1}
    \end{aligned}
\end{equation}
\ 
\begin{equation}
    \begin{aligned}
        \mathbf{e}_i^{(1)+}=\frac{1}{\sqrt{|\mathcal{N}^+_{i}|}\sqrt{|\mathcal{N}^+_{u}|}} \sum_{u\in \mathcal{N}^+_{i}} \mathbf{e}_u^{(0)}, \\
        \mathbf{e}_i^{(1)-}=\frac{1}{\sqrt{|\mathcal{N}^-_{i}|}\sqrt{|\mathcal{N}^-_{u}|}} \sum_{u\in \mathcal{N}^-_{i}} \mathbf{e}_u^{(0)}.
        \label{eq:i1}
    \end{aligned}
\end{equation}
The symmetric normalization term $\frac{1}{\sqrt{|\mathcal{N}^+_{i}|}\sqrt{|\mathcal{N}^+_{u}|}}$ and $\frac{1}{\sqrt{|\mathcal{N}^-_{i}|}\sqrt{|\mathcal{N}^-_{u}|}}$ evolve from the design of standard GCN, which can avoid the scale of embeddings increasing with graph convolution operations. $L_1$ norm and some other choices can also be applied here, while we chose this according to the performance.

\subsubsection{High-order Propagation and Layer Combination}

With the representation in the first layer, we can stack more embedding propagation layers to explore the high-order neighborhood information. Such high-order connectivities are crucial to encoding the collaborative signal to model the preferences of users and items. However, there are two obstacles that need to be avoided when processing high-order heterogeneous interactions: the assumption of balance theory no longer holds in recommender systems, and there is no homophily between nodes linked by negative edges. 

For these problems, we utilize homophily between positive neighbors to pass both positive and negative collaborative signals to capture the preferences within high-order heterogeneous interactions since users are likely to have similar negative preferences if they are similar (as shown in Fig~\ref{fig: example}).
As illustrated in Fig~\ref{fig:model}, by stacking $l$ propagation layers, a user (and an item) receives the positive and negative messages passed from its $l$-hop neighbors. The representations of user $u$ and item $i$ in the $l$-th layer can be recursively formulated as:
\begin{equation}
    \begin{aligned}
        \mathbf{e}_u^{(l+1)+}=\frac{1}{\sqrt{|\mathcal{N}^+_{u}|}\sqrt{|\mathcal{N}^+_{i}|}} \sum_{i\in \mathcal{N}^+_{u}} \mathbf{e}_i^{(l)+}, \\
        \mathbf{e}_u^{(l+1)-}=\frac{1}{\sqrt{|\mathcal{N}^+_{u}|}\sqrt{|\mathcal{N}^+_{i}|}} \sum_{i\in \mathcal{N}^+_{u}} \mathbf{e}_i^{(l)-},
        \label{eq:u2}
    \end{aligned}
\end{equation}
\begin{equation}
    \begin{aligned}
        \mathbf{e}_i^{(l+1)+}=\frac{1}{\sqrt{|\mathcal{N}^+_{i}|}\sqrt{|\mathcal{N}^+_{u}|}} \sum_{u\in \mathcal{N}^+_{i}} \mathbf{e}_u^{(l)+}, \\
        \mathbf{e}_i^{(l+1)-}=\frac{1}{\sqrt{|\mathcal{N}^+_{i}|}\sqrt{|\mathcal{N}^+_{u}|}} \sum_{u\in \mathcal{N}^+_{i}} \mathbf{e}_u^{(l)-}.
        \label{eq:i2}
    \end{aligned}
\end{equation}

In UniMo, the only trainable model parameters are the embeddings at $0$-th layer, i.e., $e_u^{(0)}$ for users and $e_i^{(0)}$ for items. The first-order and high-order positive and negative representations of users and items can be calculated via Eq.~\ref{eq:u1} - Eq.~\ref{eq:i2}. After computing the high-order preference embedding at top-$l$ layers. Similar to LightGCN, we respectively stack the positive and negative preference embeddings at each layer and take unweighted arithmetic mean to obtain the final positive and negative representations of a user (an item):
\begin{equation}
    \begin{aligned}
        \mathbf{e}_u^+&=\frac{1}{L+1}\sum_{l=0}^L \mathbf{e}_u^{(l)+};\\
        \mathbf{e}_i^+&=\frac{1}{L+1}\sum_{l=0}^L \mathbf{e}_i^{(l)+}, \\
        \mathbf{e}_u^-&=\frac{1}{L}\sum_{l=1}^L \mathbf{e}_u^{(l)-};\\
        \mathbf{e}_i^-&=\frac{1}{L}\sum_{l=1}^L \mathbf{e}_i^{(l)-},
    \end{aligned}
\end{equation}
where $\mathbf{e}_u^{(0)+}$ is equivalent to $\mathbf{e}_u^{(0)}$. It is worth noting that the arithmetic mean terms in positive and negative calculations are not completely consistent since the initial embedding contains positive preference~\cite{idsf} while negative preference is obtained from negative interactions.

Next, we introduce how to use these positive and negative embeddings to obtain recommendations.

\subsection{Negative Preference Filter and Recommendation Prediction}

To generate satisfactory recommendations for users, we calculate the positive ranking score and obtain the top$K$ items to be recommended. Then, we utilize filters to remove elements that users dislike before the final recommendation. For user $u$, the final recommendations can be expressed as:
\begin{equation}
    Rec(u)=Filter(topK(\hat{y}_{ui}^+),topK(\hat{y}_{ui}^-)),
\end{equation}
where $\hat{y}_{ui}^+=\mathbf{e}_u^{+\top}\mathbf{e}_i^+$ and $\hat{y}_{ui}^-=\mathbf{e}_u^{-\top}\mathbf{e}_i^-$ are the predicted positive and negative ranking score between user $u$ and item $i$, respectively. $Filter(\cdot)$ is the negative preference filter. Specifically, we calculate the negative ranking score and remove the negative top$K$ items from the final recommended item ranking to ensure that users won't be recommended items they dislike.

With the negative preference filter, our LSGRec takes into account the negative preference, excluding elements that users dislike, resulting in desirable recommendations.

\subsection{Multiple Learning Objectives}

To optimize the parameters in our method, we construct an end-to-end training strategy to jointly optimize the recommendation task, including positive BPR, negative BPR, rating prediction, and an orthogonality constraint.

\textbf{Positive BPR.} The original Bayesian Personalized Ranking (BPR) loss cannot reflect the difference between positive and negative interactions since it is a pairwise loss based on the relative ranking between observed and unobserved interactions by encouraging the prediction of an observed user-item pair to be higher than its unobserved counterparts. In our setting, the objective function should consider three types of relations: positive interactions, negative interactions, and unobserved interactions. Following SiReN, we propose a positive BPR loss, which encourages the predicted positive ranking score of an observed interaction to be higher than an unobserved one, along with the induced difference between positive and negative interactions:

\begin{equation}
    \mathcal{L}_{BPR}^+=-\sum_{u\in\mathcal{U}}\sum_{i\in\mathcal{N}_{u}}\sum_{j\notin\mathcal{N}_{u}}\ln\sigma(c_1\hat{y}_{ui}^+-\hat{y}_{uj}^+),
    \label{eq:bpr+}
\end{equation}
where $\sigma(\cdot)$ is the sigmoid function, $sign(\cdot)$ is the sign function, and $c_1$ is the induced term to distinguish positive and negative interactions. Through this calculation, the difference in predicted ratings between positive interactions and unobserved interactions will be greater than the difference between negative interactions and unobserved interactions.

\textbf{Negative BPR.} Similar to positive BPR, we propose negative BPR loss to constrain the proportion of negative preferences among the three types of interaction:

\begin{equation}
    \mathcal{L}_{BPR}^-=-\sum_{u\in\mathcal{U}}\sum_{i\in\mathcal{N}_{u}}\sum_{j\notin\mathcal{N}_{u}}\ln\sigma(\hat{y}_{uj}^--c_2\hat{y}_{ui}^-).
    \label{eq:bpr-}
\end{equation}
where $c_2$ is the induced term to distinguish positive and negative interactions in negative BPR loss.

\textbf{Rating Prediction.} In addition, we adopt the rating prediction to distinguish fine-grained preferences at each level, which evolves from link prediction, a widely applied task in signed graph representation learning~\cite{latentgroup}. We utilize an $MLP$ module to predict the rating of user $u$ on item $i$:
\begin{equation}
    \mathcal{L}_{MSE}=\frac{1}{|\mathcal{E}|}\left(\mbox{ReLU}([\mathbf{e}_u^+,\mathbf{e}_i^+]\mathbf{W}_{MLP}^{(1)})\mathbf{W}_{MLP}^{(2)}-\omega_{ui}\right)^2
\end{equation}
where $\mathbf{W}_{MLP}^{(1)}\in\mathbb{R}^{2d\times2d},\mathbf{W}_{MLP}^{(2)}\in\mathbb{R}^{2d\times1}$ are the trainable weight matrices of $MLP$ layers, $ReLU(\cdot)$ is the activation function.

\textbf{Orthogonality Constraint.} We apply an orthogonality constraint to force the positive and negative representation of each user and each item to share none common information:
\begin{equation}
    \mathcal{L}_{ortho}=\Vert \mathbf{e}_u^+\cdot\mathbf{e}_u^-\Vert^2 +\Vert \mathbf{e}_i^+\cdot\mathbf{e}_i^-\Vert^2
\end{equation}

\textbf{Overview.} 
We simultaneously optimize the above losses. That is, the overall objective function can be written as follows:
\begin{equation}
    \mathcal{L}=\mathcal{L}_{BPR}^++\mathcal{L}_{BPR}^-+\mathcal{L}_{MSE}+\mathcal{L}_{ortho}+\lambda\Vert\Theta\Vert_{2}^{2}
\end{equation}
where $\lambda$ and $\Theta$ represent the strengths of $L_{2}$ regularization, and the learnable parameters of the model, respectively.

Compared with other vanilla GCN algorithms, our method does not increase the time complexity of the graph convolution, although we introduce the sign of edges. At the same time, our LSGRec outperforms existing sign-aware recommendation methods in terms of not only performance but also time efficiency.

\section{Experiments and Analysis}

We conduct a series of experiments on three real-world datasets to address the following research questions:
\begin{itemize}
    \item[\textbf{RQ1:}] How does LSGRec perform compared with the state-of-the-art traditional CF methods and sign-aware recommendation methods?
    \item[\textbf{RQ2:}] How does the performance of LSGRec change when varying the number of propagation layers in the UniMo module?
    \item[\textbf{RQ3:}] What is the effectiveness of each task in the optimization and the negative preference filter?
\end{itemize}

\subsection{Experimental Settings}

In this section, we describe datasets, metrics, baselines, hyperparameters, and other details used in the experiments.

\textbf{Datasets.} 
Three public datasets are used in our experiments, including (a) \underline{Amazon-Beauty}\footnote{\url{https://cseweb.ucsd.edu/~jmcauley/datasets/amazon/links.html}}, (b) \underline{Amazon-Book}\footnotemark[1], and (c) \underline{Yelp2021}\footnote{\url{https://www.yelp.com/dataset}}. We refer to them as \emph{Beauty}, \emph{Book}, and \emph{Yelp} in brief, respectively. Table~\ref{tab:dataset} summarizes the statistics of three datasets. All datasets are split into training and testing subsets with a ratio of 8:2. For the three datasets, we use the threshold $\delta$ of 2.5 to split the original ratings as binary signals and remove users/items that have less than 5, 20, 20 interactions for Beauty, Book, Yelp respectively. Three common metrics -- $Precision@K$, Recall$@K$, and NDCG$@K$ are used to evaluate the effectiveness of our method, and $K$ is set to $10,20$ by default. We additionally calculate the average training time per epoch. 5-fold cross-validation is adopted to ensure the reliability of the experimental results.
\begin{table}[t]
    \centering
    \small
    \caption{Statistics of the datasets.  "Ratio" denotes the number ratio between positive and negative ratings in the training set.}
    \begin{tabular}{c|ccc}
        \toprule 
        \textbf{Dataset} & \textbf{Amazon-Beauty} & \textbf{Amazon-Book} & \textbf{Yelp2021} \\
        \midrule 
        {\# Users} & 22,363 & 35,736 & 41,772 \\
        {\# Items} & 12,101 & 38,121 & 30,037 \\
        {\# Interactions} & 172,188 & 1,960,674 & 2,116,215 \\
        {Density(\%)} & 0.064 & 0.14 & 0.16 \\
        {Ratio} & 1:0.13 & 1:0.07 & 1:0.16 \\
        \bottomrule
    \end{tabular}
    \label{tab:dataset}
\end{table}

\begin{table*}[!htbp]
    \centering
    \caption{Performance of all comparison methods. Each row's second-best score is underlined and the top score is highlighted in bold. The final column indicates the percentage of performance improvement relative to the second-best one. The \textit{Secs/Epoch} is the average training seconds per epoch.}
    \renewcommand\arraystretch{0.9}
    \vspace{-10pt}
    \scalebox{1}{
    \setlength\tabcolsep{1mm}{
    \begin{tabular}{c|c|p{1.4cm}<{\centering}p{1.4cm}<{\centering}p{1.4cm}<{\centering}p{1.4cm}<{\centering}p{1.8cm}<{\centering}p{1.4cm}<{\centering}|c}
    \toprule
        Dataset & {Metrix($\times100\%$)} & NGCF & LightGCN & SGCN & SiReN & PANE-GNN & \textbf{LSGRec} & Improve(\%) \\
        \midrule
        \multirow{7}{*}{Beauty} & Precision@10 & 1.286 & 1.892 & 1.149 & \underline{1.996} & 1.831 & \textbf{2.108} & 5.611  \\ 
        ~ & Recall@10 & 5.025 & 7.673 & 4.525 & \underline{8.762} & 8.017 & \textbf{9.265} & 5.741  \\ 
        ~ & NDCG@10 & 3.348 & 5.017 & 2.887 & \underline{5.802} & 5.225 & \textbf{6.143} & 5.877  \\ 
        ~ & Precision@20 & 0.990 & 1.260 & 0.921 & \underline{1.491} & 1.346 & \textbf{1.535} & 2.951  \\ 
        ~ & Recall@20 & 7.726 & 10.06 & 7.160 & \underline{12.76} & 11.59 & \textbf{13.22} & 3.605  \\ 
        ~ & NDCG@20 & 4.187 & 6.221 & 3.706 & \underline{7.036} & 6.328 & \textbf{7.365} & 4.676  \\ 
        ~ & {Secs/Epoch$\downarrow$} & 3.0 & 2.5 & 3.5 & 4.0 & 33.5 & 2.0 & - \\ 
        \midrule
        \midrule
        \multirow{7}{*}{Book} & Precision@10 & 5.499 & \underline{6.055} & 3.646 & 5.646 & 5.921 & \textbf{6.699} & 10.64  \\ 
        ~ & Recall@10 & 6.676 & 6.985 & 4.237 & 6.858 & \underline{7.025} & \textbf{8.164} & 16.21  \\ 
        ~ & NDCG@10 & 7.937 & 8.012 & 4.822 & 7.755 & \underline{8.018} & \textbf{9.269} & 15.60  \\ 
        ~ & Precision@20 & 4.275 & \underline{4.842} & 3.104 & 4.588 & 4.819 & \textbf{5.367} & 10.84  \\ 
        ~ & Recall@20 & 10.63 & 11.05 & 7.069 & 10.81 & \underline{11.14} & \textbf{12.63} & 13.38  \\ 
        ~ & NDCG@20 & 8.783 & \underline{9.294} & 5.736 & 8.974 & 9.261 & \textbf{10.62} & 14.27  \\ 
        ~ & {Secs/Epoch$\downarrow$} & 39.5 & 34.0 & 40.0 & 83.5 & 739.5 & 38.5 & - \\ 
        \midrule
        \midrule
        \multirow{7}{*}{Yelp} & Precision@10 & 2.721 & 2.919 & 2.677 & \underline{4.213} & 4.059 & \textbf{4.411} & 4.699 \\ 
        ~ & Recall@10 & 4.033 & 4.234 & 3.336 & \underline{5.304} & 5.098 & \textbf{5.516} & 3.997  \\ 
        ~ & NDCG@10 & 3.872 & 4.308 & 3.705 & \underline{5.787} & 5.643 & \textbf{6.107} & 5.529  \\ 
        ~ & Precision@20 & 2.566 & 2.614 & 2.234 & \underline{3.487} & 3.359 & \textbf{3.641} & 4.416  \\ 
        ~ & Recall@20 & 5.825 & 6.682 & 5.512 & \underline{8.726} & 8.282 & \textbf{8.996} & 3.094  \\ 
        ~ & NDCG@20 & 4.522 & 4.833 & 4.388 & \underline{6.908} & 6.643 & \textbf{7.188} & 4.053  \\ 
        ~ & {Secs/Epoch$\downarrow$} & 65.0 & 41.5 & 76.5 & 89.5 & 892.0 & 69.0 & - \\ 
        \bottomrule
    \end{tabular}
    }}
    \label{tab:overall}
\end{table*}

\textbf{Baselines.}
We compare LSGRec with five competing methods, including two unsigned GCN-based CF methods, a standard Signed Graph Convolutional Network (SGCN), and two state-of-the-art methods in the sign-aware recommendation.
\begin{itemize}
    \item \textbf{NGCF}~\cite{ngcf} explicitly integrates the user-item interactions into the embedding process, learning the topology by graph convolution, and effectively harvests the high-order collaborative signals for recommendation.
    \item \textbf{LightGCN}~\cite{lightgcn} abandons the feature transformation and nonlinear activation in standard GCN, only retaining the neighbor aggregation and layer combination for collaborative filtering.
    \item \textbf{SGCN}~\cite{sgcn} utilizes balance theory to aggregate and propagate the information, modeling positive and negative edges in a whole signed graph.
    \item \textbf{SiReN}~\cite{siren} constructs a signed bipartite graph for modeling users' preferences and generates positive and negative embeddings for the partitioned graphs. In addition, SiReN designs a sign-aware Bayesian Personalized Ranking (BPR) loss function to induce differences between high and low ratings.
    \item \textbf{PANE-GNN}~\cite{pane-gnn} is one of the state-of-the-art sign-aware recommendation methods, which incorporates users' positive and negative preferences by distinct message-passing mechanisms and contrastive learning.
\end{itemize}
We do not include some methods in signed graph neural networks like GSGNN~\cite{latentgroup}, SBGCL~\cite{sbgcl}, and SIGAT~\cite{sigat} since they are designed for link prediction. LightSGCN~\cite{lightgcn,negpos} does not provide open-source codes for reliable reproduction for recommendations.

\textbf{Hyperparameters.} For a fair comparison, we set the embedding dimension to 64, batch size to 1024, and initialized all model parameters with Xavier initializer~\cite{Xavier}, which is optimized by Adam optimizer~\cite{AdamAM}. The learning rate is set to $0.005$, and \textit{MultiStepLR} is utilized to schedule the learning rate. We test the number of propagation layers $L$ in $[1,2,3,4]$. Besides, we train all methods for $200$ epochs and record the best performance according to Recall$@10$. For all baselines, we follow the original settings of comparison methods to achieve the best performance.

\subsection{Performance Comparison (RQ1)}

The comparative results are summarized in Table~\ref{tab:overall}, from which we can find that our proposed LSGRec outperforms both unsigned and signed baselines. Generally, sign-aware recommendation methods perform better than unsigned methods, implying the value of modeling positive and negative user preferences from the sign of interactions.

SGCN obtains the worst performance due to its dependency on the balance theory. It introduces noisy or incorrect collaborative information during convolution since the balance theory mistakenly treats some nodes as positive neighbors ("the enemy of my enemy"). This finding suggests that the assumption of balance theory, which is designed for signed unipartite graphs, is not applicable to recommendation scenarios where users typically have diverse interests.

For unsigned recommendation methods, LightGCN performs better than NGCF thanks to its abandonment of feature transformation and nonlinear activation, which are burdensome for capturing collaborative signals in graphs, and the proposal of an optimized lightweight aggregation method (rather than the aggregation in NGCF). The optimized aggregation captures more precise high-order preferences that can help improve recommendation performance.

The sign-aware baselines overall outperform the unsigned baselines on \emph{Beauty} and \emph{Yelp} since they can more precisely represent users' preferences and capture more detailed collaborative signals from the sign of edges with less information loss. In comparison to PANE-GNN, SiReN surpasses it on these two datasets because PANE-GNN adopts graph convolution on the negative graph, while negative edges cannot convey high-order similarity~\cite{ghcf,sigat}. In other words, we cannot assume that there are collaborative signals between nodes linked by multi-hop negative edges. SiReN utilizes an MLP to encode the negative graph, avoiding these underlying noise.

It is worth noting that the performances of unsigned methods on \emph{Book} are comparable to the sign-aware methods. This is because the ratio of negative interactions in this dataset is lower than in others, as shown in Table~\ref{tab:dataset}. If independent modeling approaches are adopted, high-order heterogeneous interactions will be lost after separating the whole graph, and the negative graph will be a forest instead of a connected graph, resulting in difficulty for negative encoders to capture effective negative preferences.
That is, existing sign-aware methods heavily depend on adequate negative interactions, which may not be met in all real-world recommendation scenarios. Meanwhile, due to the adoption of contrastive learning, PANE-GNN is more robust and performs better than SiReN.

\begin{table*}[t]
    \centering
    \caption{Performance of all variants of our LSGRec.}
    \vspace{-10pt}
    \scalebox{1}{
    \setlength\tabcolsep{1mm}{
    \begin{tabular}{c|c|cccccc}
    \toprule
        Dataset & {Variant} & Precision@10 & Recall@10 & NDCG@10 & Precision@20 & Recall@20 & NDCG@20 \\
        \midrule
        \multirow{4}{*}{Beauty} & \textbf{LSGRec} & 2.11 & 9.26 & 6.14 & 1.53 & 13.22 & 7.36 \\
        ~ & \ \textit{w/o} $\mathcal{L}_{BPR}^-$ & 2.09 & 9.23 & 6.09 & 1.53 & 13.22 & 7.31 \\
        ~ & \ \textit{w/o} $\mathcal{L}_{MSE}$ & 2.09 & 9.21 & 6.06 & 1.53 & 13.18 & 7.28 \\ 
        ~ & \ \textit{w/o} $\mathcal{L}_{ortho}$ & 2.07 & 9.09 & 6.04 & 1.52 & 13.19 & 7.29 \\ 
        ~ & \ \textit{w/o filter} & 2.08 & 9.17 & 6.05 & 1.52 & 13.19 & 7.28 \\ 
        \midrule
        \midrule
        \multirow{4}{*}{Book} & \textbf{LSGRec} & 6.69 & 8.16 & 9.26 & 5.36 & 12.63 & 10.61 \\
        ~ & \ \textit{w/o} $\mathcal{L}_{BPR}^-$ & 6.41 & 7.88 & 8.91 & 5.17 & 12.31 & 10.28 \\ 
        ~ & \ \textit{w/o} $\mathcal{L}_{MSE}$ & 6.41 & 7.86 & 8.88 & 5.16 & 12.27 & 10.23 \\ 
        ~ & \ \textit{w/o} $\mathcal{L}_{ortho}$ & 6.43 & 7.89 & 8.93 & 5.17 & 12.32 & 10.31 \\ 
        ~ & \ \textit{w/o filter} & 6.45 & 7.88 & 8.93 & 5.19 & 12.29 & 10.28 \\ 
        \midrule
        \midrule
        \multirow{4}{*}{Yelp} & \textbf{LSGRec} & 4.41 & 5.51 & 6.11 & 3.64 & 8.99 & 7.18 \\
        ~ & \ \textit{w/o} $\mathcal{L}_{BPR}^-$ & 4.15 & 5.17 & 5.72 & 3.43 & 8.47 & 6.74 \\ 
        ~ & \ \textit{w/o} $\mathcal{L}_{MSE}$ & 4.34 & 5.39 & 5.99 & 3.57 & 8.84 & 7.05 \\ 
        ~ & \ \textit{w/o} $\mathcal{L}_{ortho}$ & 4.05 & 5.01 & 5.55 & 3.35 & 8.29 & 6.56 \\ 
        ~ & \ \textit{w/o filter} & 4.09 & 5.04 & 5.61 & 3.38 & 8.31 & 6.62 \\ 
        \bottomrule
    \end{tabular}
    }}
    \label{tab:variants}
\end{table*}

Finally, our proposed LSGRec beats the second-best baselines in terms of three metrics by around 2.9-5.8\% on \emph{Beauty}, 10.6-14.2\% on \emph{Book}, 3.1-5.5\% on \emph{Yelp}, respectively. We attribute these significant improvements mainly to learning the complete negative preferences from direct neighbors and high-order heterogeneous interactions.
The comprehensive employment of positive and negative interactions in the signed graph provides valuable positive and negative preferences. Meanwhile, we propose a negative preference filter which can help ensure satisfactory recommendation results. On \emph{Book} dataset, LSGRec outperforms all baselines by a large margin, indicating that even with a small number of negative interactions, our unified modeling method captures correct and precise high-order collaborative signals, improving the recommendations by utilizing comprehensive representations of users and items.

\subsection{Hyperparameter Analysis (RQ2)}

\begin{figure}[htbp]
    \centering
    \includegraphics[width=\linewidth]{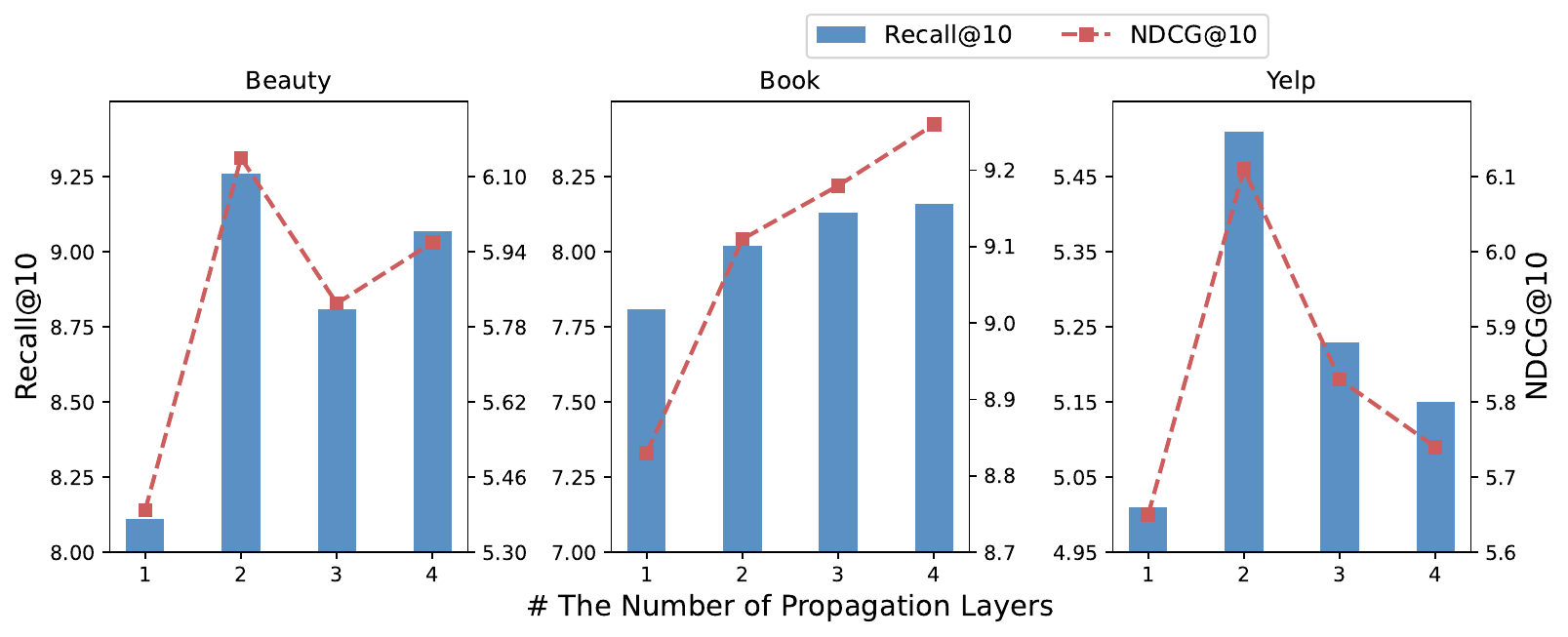}
    \caption{Recall$@10$ and NDCG$@10$ across various numbers of propagation layers (i.e., the value of $L$).}
    \label{fig:layers}
\end{figure}

We examine the performance of our LSGRec with various numbers of propagation (i.e., the value of $L$) in the range of $\{1,2,3,4\}$. Fig.~\ref{fig:layers} reports the results of the performance comparison. The main observations are as follows:
\begin{itemize}
    \item The performance on all datasets reaches its lowest value when $L$ is set to 1, which considers the first-order neighbors only, indicating that higher-order collaborative signals are necessary for capturing user preferences.
    \item When further stacking the propagation layer, we find that UniMo leads to overfitting on \emph{Beauty} and \emph{Yelp} datasets. This might be caused by applying a too-deep architecture, resulting in over-smoothing. The marginal improvements on these two datasets verify that conducting two propagation layers is sufficient to capture the collaborative signals.
    \item On \emph{Book} dataset, the performance improves as the number of layers increases. Since the ratio of negative interactions is lower, increasing the number of layers will significantly improve the user's negative preferences.
    \item Increasing the depth of UniMo enhances the recommendation cases. When $L$ is adjusted to 2, 4, and 2, respectively, performance on \emph{Beauty}, \emph{Book}, \emph{Yelp} yields the best results. Appropriate $L$ can obtain better representations of users and items by aggregating collaborative signals. 
\end{itemize}

\subsection{Ablation Study (RQ3)}

\begin{figure}[htbp]
    \centering
    \includegraphics[width=\linewidth]{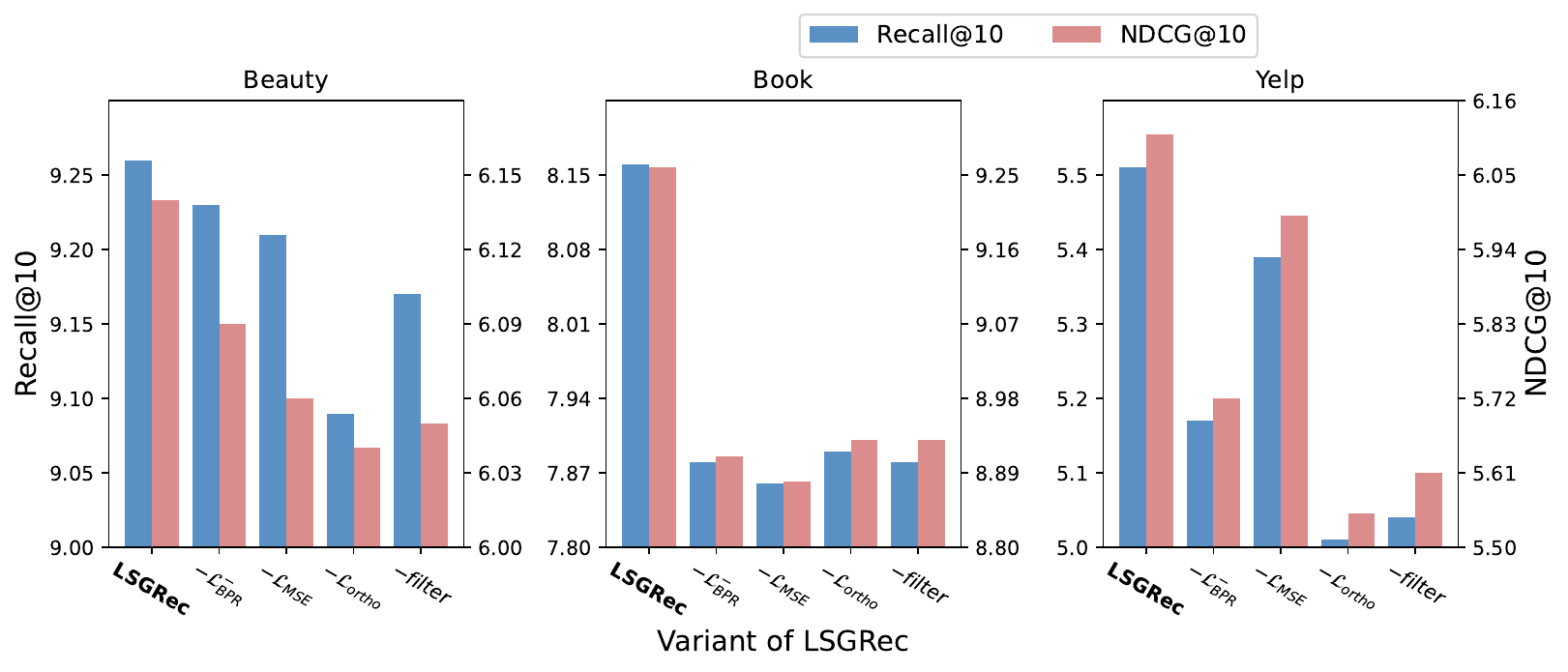}
    \caption{Recall$@10$ and NDCG$@10$ across different variants of our LSGRec. '-' represents '\textit{w/o}' for brief.}
    \label{fig:variants}
\end{figure}

To explore the effects of each auxiliary task and the negative preference filter, we compare the results on four variants: \textit{w/o} $\mathcal{L}_{BPR}^-$, which discards negative BPR task, \textit{w/o} $\mathcal{L}_{MSE}$ which ignore the rating prediction task, \textit{w/o} $\mathcal{L}_{ortho}$, which leaves out orthogonality constraint when training and \textit{w/o filter} which skips negative preference filtering when generating recommendations. Table~\ref{tab:variants} summarizes the performance of different variants of LSGRec, and Fig.~\ref{fig:variants} intuitively presents Recall$@10$ and NDCG$@10$ change in different variants, from which we have the following observations.

Without the negative BPR task, the performance of LSGRec \textit{w/o} $\mathcal{L}_{BPR}^-$ declines on all three datasets, especially on \emph{Book} and \emph{Yelp}, indicating that controlling the proportion of negative preference among different types of interactions is essential, which can train representations of users and items to precisely capture and model different interactions as an aid to positive BPR task. When discarding the rating prediction task, the performance of LSGRec \textit{w/o} $\mathcal{L}_{MSE}$ drops on all datasets. It helps the model capture the more fine-grained level of interactions and improve the representations' perception of each score on top of the two BPR tasks. Based on these findings, we observe that the above two tasks cooperate with the positive BPR task, more accurately modeling the preferences of users and items. In addition, we find that the orthogonality constraint has almost the largest impact on the performance among these tasks (LSGRec \textit{w/o} $\mathcal{L}_{ortho}$ has a more significant performance decline). This is because, without the constraint, positive and negative representations may be mixed since they are learned from the same initial embedding. The orthogonality constraint can force the decoupling of positive and negative embeddings during graph convolution.

Besides these auxiliary tasks, it is necessary to draw the negative preference filter into generating recommendations. Due to the lack of the negative preference filter, the performance of LSGRec \textit{w/o filter} decreases compared to the original model. We believe not filtering before the recommendation may lead to recommending items that users dislike. For example, for a user who likes sports but is disgusted with soccer, the recommender system should filter soccer-related items out before predicting.

Moreover, we observe that the decrease in performance is smaller on the \emph{Beauty}. We believe this is because the \emph{Beauty} has fewer interactions and higher sparsity, so it is not as sensitive to the refinement of interactions as others.

Finally, LSGRec outperforms all variants on three datasets across three evaluation metrics, further validating the significance of leveraging fine-grained interactions to capture accurate preferences with these auxiliary tasks.

\section{Related Work}

\subsection{Graph-based Recommendation}

Using graph neural networks (GNNs)~\cite{scarselli2008graph, wu2022graph} in modeling interactions between users and items has gained broad acknowledgment as potent architectures. Numerous recommendation models, built upon GNNs as their foundations, prove that they have attained state-of-the-art performance across various sub-fields~\cite{li2023graph, chen2020revisiting, yu2020enhancing, lightgcn, fan2019graph}. For example, NGCF~\cite{ngcf} leveraged graph convolutional networks (GCNs) to transform the interaction graph into latent embeddings. GraphRec~\cite{fan2019graph} provided a principled approach to capture interactions and opinions in the user-item graph jointly. To simplify the graph message passing algorithm, LightGCN~\cite{lightgcn} removed the redundant operations, including transformation matrices and nonlinear activation functions in NGCF, achieving better results in recommendation performance and efficiency. SGL~\cite{wu2021self} adopts self-supervised learning to achieve more accurate user and item representations. However, negative interactions are overlooked in most cases, which leads to inaccurate user preferences.

It is worth mentioning that several recent efforts have paid attention to better modeling users/items with positive and negative edges in signed user-item bipartite graphs. Specifically, SiReN~\cite{siren} constructs a signed bipartite graph and generates positive and negative embeddings for the partitioned graphs. PANE-GNN~\cite{pane-gnn} employs contrastive learning on the negative graph to reduce noise and filter out items with high disinterest scores, ensuring the relevance of the recommended results. SiGRec~\cite{negpos} investigates three kinds of negative feedback and defines a new sign cosine loss function to adaptively capture differences among them. Dual-LightGCN~\cite{Huang2023DualLightGCN} divides the original user–item interaction graph into two bipartite subgraphs and performs the LightGCN model on them. One subgraph is used to model the preferences between users and items, while the other is used to model the dislike relationships between them. 
Nevertheless, these methods adopt independent encoders for modeling positive and negative interactions, which may corrupt high-order collaborative information in the signed graph.

\subsection{Signed Graph Neural Network}

Signed graphs have been widely explored in social networks, focusing on node-level and graph-level tasks, e.g., node classification~\cite{signet}, signed link prediction~\cite{latentgroup,sine,sigat,snea,sbgnn,sgcl,sgcn}, graph classification~\cite{sbgcl}, node ranking~\cite{beside}, etc. SiNE~\cite{sine} first introduces the balance theory into the loss function. Then SGCN~\cite{sgcn} integrates the balance theory into graph convolution operations and becomes the mainstream paradigm followed by subsequent works. Similarly, some approaches~\cite{sigat,snea,sdgnn} utilize graph attention~\cite{gat} and transformer~\cite{graphtransformer} architectures to build graph neural networks and learn signed graph representations. These methods for signed graphs are built upon the assumption of balance theory~\cite{Cartwright1977STRUCTURALBA}, which implies that "the friend of my friend tends to be my friend" and "the enemy of my enemy tends to be my friend". However, "the enemy of my enemy tends to my friend" no longer holds in the context of recommendation~\cite{negpos,pane-gnn,siren,sbgnn,Xu2023SignedNetwork,Xu2022Dualbranch} so that these methods cannot be applied to the field of recommendations.

\section{Conclusion and Future Work}

In this paper, we explored a fundamentally significant problem of how to fully use different types of user-item interactions in developing graph-based recommendation methods. We first revisited the shortcomings in previous sign-aware recommender systems. They adopt independent encoders for each type of interaction, disrupting high-order collaborative signals in the signed graph and overlooking the negative preferences within high-order heterogeneous interactions. Then, we analyzed the challenges in adapting existing signed graph neural networks to graph-based recommender systems. These methods for unipartite signed graphs are built upon the assumption of balance theory, which no longer holds in the context of recommendation. To tackle these issues, we proposed a novel model, termed LSGRec, which adopts a unified modeling approach to simultaneously model users' positive and negative preferences on a bipartite signed user-item interaction graph based on the homophily instead of the balance theory and optimizes them with a multi-task training strategy. We also introduced a negative preference filter to generate revised recommendations by filtering items that users dislike. Experiments on three real-world datasets demonstrated that our LSGRec method remarkably outperforms two state-of-the-art sign-aware recommendation methods and two baseline graph-based methods. 

Our work provides a new insight into designing a new graph neural network on a whole graph to model detailed user preferences (e.g., like or dislike). To highlight the conceptual design, we have simplified many model architectures as much as possible, such as the difference between positive and negative samples in the sign-aware BPR loss functions, normalization terms in convolution, and adaptive weights in layer combination. We leave these more detailed parameters and model designs for future work.

\begin{acks}
To Robert, for the bagels and explaining CMYK and color spaces.
\end{acks}

\bibliographystyle{ACM-Reference-Format}
\bibliography{reference}


\begin{thebibliography}{45}


\ifx \showCODEN    \undefined \def \showCODEN     #1{\unskip}     \fi
\ifx \showDOI      \undefined \def \showDOI       #1{#1}\fi
\ifx \showISBNx    \undefined \def \showISBNx     #1{\unskip}     \fi
\ifx \showISBNxiii \undefined \def \showISBNxiii  #1{\unskip}     \fi
\ifx \showISSN     \undefined \def \showISSN      #1{\unskip}     \fi
\ifx \showLCCN     \undefined \def \showLCCN      #1{\unskip}     \fi
\ifx \shownote     \undefined \def \shownote      #1{#1}          \fi
\ifx \showarticletitle \undefined \def \showarticletitle #1{#1}   \fi
\ifx \showURL      \undefined \def \showURL       {\relax}        \fi
\providecommand\bibfield[2]{#2}
\providecommand\bibinfo[2]{#2}
\providecommand\natexlab[1]{#1}
\providecommand\showeprint[2][]{arXiv:#2}

\bibitem[Cai et~al\mbox{.}(2022)]%
        {CAI2022102917}
\bibfield{author}{\bibinfo{person}{Shensheng Cai}, \bibinfo{person}{Wei Shan}, {and} \bibinfo{person}{Mingli Zhang}.} \bibinfo{year}{2022}\natexlab{}.
\newblock \showarticletitle{Structure information learning for neutral links in signed network embedding}.
\newblock \bibinfo{journal}{\emph{Information Processing \& Management}} (\bibinfo{year}{2022}), \bibinfo{pages}{102917}.
\newblock


\bibitem[Cartwright and Harary(1977)]%
        {Cartwright1977STRUCTURALBA}
\bibfield{author}{\bibinfo{person}{Dorwin Cartwright} {and} \bibinfo{person}{Frank Harary}.} \bibinfo{year}{1977}\natexlab{}.
\newblock \showarticletitle{STRUCTURAL BALANCE: A GENERALIZATION OF HEIDER'S THEORY1}. In \bibinfo{booktitle}{\emph{Psychological Review}}. \bibinfo{pages}{277}.
\newblock


\bibitem[Chen et~al\mbox{.}(2021)]%
        {ghcf}
\bibfield{author}{\bibinfo{person}{Chong Chen}, \bibinfo{person}{Weizhi Ma}, \bibinfo{person}{Min Zhang}, \bibinfo{person}{Zhaowei Wang}, \bibinfo{person}{Xiuqiang He}, \bibinfo{person}{Chenyang Wang}, \bibinfo{person}{Yiqun Liu}, {and} \bibinfo{person}{Shaoping Ma}.} \bibinfo{year}{2021}\natexlab{}.
\newblock \showarticletitle{Graph Heterogeneous Multi-Relational Recommendation}. In \bibinfo{booktitle}{\emph{AAAI Conference on Artificial Intelligence}}. \bibinfo{pages}{3958--3966}.
\newblock


\bibitem[Chen et~al\mbox{.}(2020)]%
        {chen2020revisiting}
\bibfield{author}{\bibinfo{person}{Lei Chen}, \bibinfo{person}{Le Wu}, \bibinfo{person}{Richang Hong}, \bibinfo{person}{Kun Zhang}, {and} \bibinfo{person}{Meng Wang}.} \bibinfo{year}{2020}\natexlab{}.
\newblock \showarticletitle{Revisiting graph based collaborative filtering: A linear residual graph convolutional network approach}. In \bibinfo{booktitle}{\emph{AAAI Conference on Artificial Intelligence}}. \bibinfo{pages}{27--34}.
\newblock


\bibitem[Chen et~al\mbox{.}(2018)]%
        {beside}
\bibfield{author}{\bibinfo{person}{Yiqi Chen}, \bibinfo{person}{Tieyun Qian}, \bibinfo{person}{Huan Liu}, {and} \bibinfo{person}{K. Sun}.} \bibinfo{year}{2018}\natexlab{}.
\newblock \showarticletitle{"Bridge": Enhanced Signed Directed Network Embedding}.
\newblock \bibinfo{journal}{\emph{Proceedings of the 27th ACM International Conference on Information and Knowledge Management}} (\bibinfo{year}{2018}), \bibinfo{pages}{773–782}.
\newblock


\bibitem[Defferrard et~al\mbox{.}(2016)]%
        {Defferrard2016ConvolutionalNN}
\bibfield{author}{\bibinfo{person}{Micha{\"e}l Defferrard}, \bibinfo{person}{Xavier Bresson}, {and} \bibinfo{person}{Pierre Vandergheynst}.} \bibinfo{year}{2016}\natexlab{}.
\newblock \showarticletitle{Convolutional Neural Networks on Graphs with Fast Localized Spectral Filtering}. In \bibinfo{booktitle}{\emph{Advances in Neural Information Processing Systems 34}}. \bibinfo{pages}{1--9}.
\newblock


\bibitem[Derr et~al\mbox{.}(2018)]%
        {sgcn}
\bibfield{author}{\bibinfo{person}{Tyler Derr}, \bibinfo{person}{Yao Ma}, {and} \bibinfo{person}{Jiliang Tang}.} \bibinfo{year}{2018}\natexlab{}.
\newblock \showarticletitle{Signed Graph Convolutional Networks}. In \bibinfo{booktitle}{\emph{2018 IEEE International Conference on Data Mining (ICDM)}}. \bibinfo{pages}{929--934}.
\newblock


\bibitem[Fan et~al\mbox{.}(2019a)]%
        {Fan2019GraphNN}
\bibfield{author}{\bibinfo{person}{Wenqi Fan}, \bibinfo{person}{Yao Ma}, \bibinfo{person}{Qing Li}, \bibinfo{person}{Yuan He}, \bibinfo{person}{Eric Zhao}, \bibinfo{person}{Jiliang Tang}, {and} \bibinfo{person}{Dawei Yin}.} \bibinfo{year}{2019}\natexlab{a}.
\newblock \showarticletitle{Graph Neural Networks for Social Recommendation}. In \bibinfo{booktitle}{\emph{The World Wide Web Conference}}. \bibinfo{pages}{417–426}.
\newblock


\bibitem[Fan et~al\mbox{.}(2019b)]%
        {fan2019graph}
\bibfield{author}{\bibinfo{person}{Wenqi Fan}, \bibinfo{person}{Yao Ma}, \bibinfo{person}{Qing Li}, \bibinfo{person}{Yuan He}, \bibinfo{person}{Eric Zhao}, \bibinfo{person}{Jiliang Tang}, {and} \bibinfo{person}{Dawei Yin}.} \bibinfo{year}{2019}\natexlab{b}.
\newblock \showarticletitle{Graph neural networks for social recommendation}. In \bibinfo{booktitle}{\emph{The World Wide Web Conference}}. \bibinfo{pages}{417--426}.
\newblock


\bibitem[Franco et~al\mbox{.}(2008)]%
        {scarselli2008graph}
\bibfield{author}{\bibinfo{person}{Scarselli Franco}, \bibinfo{person}{Gori Marco}, \bibinfo{person}{Tsoi~Ah Chung}, \bibinfo{person}{Hagenbuchner Markus}, {and} \bibinfo{person}{Monfardini Gabriele}.} \bibinfo{year}{2008}\natexlab{}.
\newblock \showarticletitle{The graph neural network model}.
\newblock \bibinfo{journal}{\emph{IEEE Transactions on neural networks}} (\bibinfo{year}{2008}), \bibinfo{pages}{61--80}.
\newblock


\bibitem[Glorot and Bengio(2010)]%
        {Xavier}
\bibfield{author}{\bibinfo{person}{Xavier Glorot} {and} \bibinfo{person}{Yoshua Bengio}.} \bibinfo{year}{2010}\natexlab{}.
\newblock \showarticletitle{Understanding the difficulty of training deep feedforward neural networks}. In \bibinfo{booktitle}{\emph{Artificial Intelligence and Statistics}}. \bibinfo{pages}{1--8}.
\newblock


\bibitem[Gori and Pucci(2007)]%
        {Gori2007ItemRankAR}
\bibfield{author}{\bibinfo{person}{Marco Gori} {and} \bibinfo{person}{Augusto Pucci}.} \bibinfo{year}{2007}\natexlab{}.
\newblock \showarticletitle{ItemRank: A Random-Walk Based Scoring Algorithm for Recommender Engines}. In \bibinfo{booktitle}{\emph{International Joint Conference on Artificial Intelligence}}. \bibinfo{pages}{2766--2771}.
\newblock


\bibitem[Hamilton et~al\mbox{.}(2017)]%
        {graphsage}
\bibfield{author}{\bibinfo{person}{William~L. Hamilton}, \bibinfo{person}{Zhitao Ying}, {and} \bibinfo{person}{Jure Leskovec}.} \bibinfo{year}{2017}\natexlab{}.
\newblock \showarticletitle{Inductive Representation Learning on Large Graphs}. In \bibinfo{booktitle}{\emph{Neural Information Processing Systems}}. \bibinfo{pages}{1025--1035}.
\newblock


\bibitem[He et~al\mbox{.}(2020)]%
        {lightgcn}
\bibfield{author}{\bibinfo{person}{Xiangnan He}, \bibinfo{person}{Kuan Deng}, \bibinfo{person}{Xiang Wang}, \bibinfo{person}{Yan Li}, \bibinfo{person}{Yongdong Zhang}, {and} \bibinfo{person}{Meng Wang}.} \bibinfo{year}{2020}\natexlab{}.
\newblock \showarticletitle{LightGCN: Simplifying and Powering Graph Convolution Network for Recommendation}. In \bibinfo{booktitle}{\emph{Proceedings of the 43rd International ACM SIGIR Conference on Research and Development in Information Retrieval}}. \bibinfo{pages}{639–648}.
\newblock


\bibitem[Huang et~al\mbox{.}(2021a)]%
        {sbgnn}
\bibfield{author}{\bibinfo{person}{Junjie Huang}, \bibinfo{person}{Huawei Shen}, \bibinfo{person}{Qi Cao}, \bibinfo{person}{Shuchang Tao}, {and} \bibinfo{person}{Xueqi Cheng}.} \bibinfo{year}{2021}\natexlab{a}.
\newblock \showarticletitle{Signed Bipartite Graph Neural Networks}.
\newblock \bibinfo{journal}{\emph{Proceedings of the 30th ACM International Conference on Information \& Knowledge Management}} (\bibinfo{year}{2021}), \bibinfo{pages}{740–749}.
\newblock


\bibitem[Huang et~al\mbox{.}(2021b)]%
        {sdgnn}
\bibfield{author}{\bibinfo{person}{Junjie Huang}, \bibinfo{person}{Huawei Shen}, \bibinfo{person}{Liang Hou}, {and} \bibinfo{person}{Xueqi Cheng}.} \bibinfo{year}{2021}\natexlab{b}.
\newblock \showarticletitle{SDGNN: Learning Node Representation for Signed Directed Networks}. In \bibinfo{booktitle}{\emph{AAAI Conference on Artificial Intelligence}}. \bibinfo{pages}{196--203}.
\newblock


\bibitem[Huang et~al\mbox{.}(2023b)]%
        {negpos}
\bibfield{author}{\bibinfo{person}{Junjie Huang}, \bibinfo{person}{Ruobing Xie}, \bibinfo{person}{Qi Cao}, \bibinfo{person}{Huawei Shen}, \bibinfo{person}{Shaoliang Zhang}, \bibinfo{person}{Feng Xia}, {and} \bibinfo{person}{Xueqi Cheng}.} \bibinfo{year}{2023}\natexlab{b}.
\newblock \showarticletitle{Negative Can Be Positive: Signed Graph Neural Networks for Recommendation}.
\newblock \bibinfo{journal}{\emph{Information Processing \& Management}} (\bibinfo{year}{2023}), \bibinfo{pages}{103403--103417}.
\newblock


\bibitem[Huang and Hou(2019)]%
        {sigat}
\bibfield{author}{\bibinfo{person}{Shen~Huawei Huang, Junjie} {and} \bibinfo{person}{Cheng~Xueqi Hou, Liang}.} \bibinfo{year}{2019}\natexlab{}.
\newblock \showarticletitle{Signed Graph Attention Networks}. In \bibinfo{booktitle}{\emph{Artificial Neural Networks and Machine Learning -- ICANN 2019: Workshop and Special Sessions}}. \bibinfo{pages}{566--577}.
\newblock


\bibitem[Huang et~al\mbox{.}(2023a)]%
        {Huang2023DualLightGCN}
\bibfield{author}{\bibinfo{person}{Wenqing Huang}, \bibinfo{person}{Fei Hao}, \bibinfo{person}{Jiaxing Shang}, \bibinfo{person}{Wangyang Yu}, \bibinfo{person}{Shengke Zeng}, \bibinfo{person}{Carmen Bisogni}, {and} \bibinfo{person}{Vincenzo Loia}.} \bibinfo{year}{2023}\natexlab{a}.
\newblock \showarticletitle{Dual-LightGCN: Dual light graph convolutional network for discriminative recommendation}.
\newblock \bibinfo{journal}{\emph{Comput. Commun.}} (\bibinfo{year}{2023}), \bibinfo{pages}{89--100}.
\newblock


\bibitem[Islam et~al\mbox{.}(2017)]%
        {signet}
\bibfield{author}{\bibinfo{person}{Mohammad~Raihanul Islam}, \bibinfo{person}{B.~Aditya Prakash}, {and} \bibinfo{person}{Naren Ramakrishnan}.} \bibinfo{year}{2017}\natexlab{}.
\newblock \showarticletitle{SIGNet: Scalable Embeddings for Signed Networks}. In \bibinfo{booktitle}{\emph{Pacific-Asia Conference on Knowledge Discovery and Data Mining}}. \bibinfo{pages}{42--54}.
\newblock


\bibitem[Jeunen(2019)]%
        {Olivier2019Revisiting}
\bibfield{author}{\bibinfo{person}{Olivier Jeunen}.} \bibinfo{year}{2019}\natexlab{}.
\newblock \showarticletitle{Revisiting offline evaluation for implicit-feedback recommender systems}. In \bibinfo{booktitle}{\emph{Proceedings of the 13th ACM Conference on Recommender Systems}}. \bibinfo{pages}{596–600}.
\newblock


\bibitem[Kingma and Ba(2015)]%
        {AdamAM}
\bibfield{author}{\bibinfo{person}{Diederik~P. Kingma} {and} \bibinfo{person}{Jimmy Ba}.} \bibinfo{year}{2015}\natexlab{}.
\newblock \showarticletitle{Adam: A Method for Stochastic Optimization}.
\newblock \bibinfo{journal}{\emph{CoRR}} (\bibinfo{year}{2015}).
\newblock


\bibitem[Kipf and Welling(2016)]%
        {gcn1st}
\bibfield{author}{\bibinfo{person}{Thomas Kipf} {and} \bibinfo{person}{Max Welling}.} \bibinfo{year}{2016}\natexlab{}.
\newblock \showarticletitle{Semi-Supervised Classification with Graph Convolutional Networks}.
\newblock \bibinfo{journal}{\emph{ArXiv}} (\bibinfo{year}{2016}), \bibinfo{pages}{1--14}.
\newblock


\bibitem[Li et~al\mbox{.}(2023)]%
        {li2023graph}
\bibfield{author}{\bibinfo{person}{Chaoliu Li}, \bibinfo{person}{Lianghao Xia}, \bibinfo{person}{Xubin Ren}, \bibinfo{person}{Yaowen Ye}, \bibinfo{person}{Yong Xu}, {and} \bibinfo{person}{Chao Huang}.} \bibinfo{year}{2023}\natexlab{}.
\newblock \showarticletitle{Graph Transformer for Recommendation}. In \bibinfo{booktitle}{\emph{Proceedings of the 46th International ACM SIGIR}}. \bibinfo{pages}{1680–1689}.
\newblock


\bibitem[Li et~al\mbox{.}(2020)]%
        {snea}
\bibfield{author}{\bibinfo{person}{Yu Li}, \bibinfo{person}{Yuan Tian}, \bibinfo{person}{Jiawei Zhang}, {and} \bibinfo{person}{Yi Chang}.} \bibinfo{year}{2020}\natexlab{}.
\newblock \showarticletitle{Learning Signed Network Embedding via Graph Attention}. In \bibinfo{booktitle}{\emph{AAAI Conference on Artificial Intelligence}}. \bibinfo{pages}{4772--4779}.
\newblock


\bibitem[Liu et~al\mbox{.}(2021)]%
        {latentgroup}
\bibfield{author}{\bibinfo{person}{Haoxin Liu}, \bibinfo{person}{Ziwei Zhang}, \bibinfo{person}{Peng Cui}, \bibinfo{person}{Yafeng Zhang}, \bibinfo{person}{Qiang Cui}, \bibinfo{person}{Jiashuo Liu}, {and} \bibinfo{person}{Wenwu Zhu}.} \bibinfo{year}{2021}\natexlab{}.
\newblock \showarticletitle{Signed Graph Neural Network with Latent Groups}.
\newblock \bibinfo{journal}{\emph{Proceedings of the 27th ACM SIGKDD Conference on Knowledge Discovery \& Data Mining}} (\bibinfo{year}{2021}), \bibinfo{pages}{1066–1075}.
\newblock


\bibitem[Liu et~al\mbox{.}(2023b)]%
        {idsf}
\bibfield{author}{\bibinfo{person}{Yuting Liu}, \bibinfo{person}{Enneng Yang}, \bibinfo{person}{Yizhou Dang}, \bibinfo{person}{Guibing Guo}, \bibinfo{person}{Qiang Liu}, \bibinfo{person}{Yuliang Liang}, \bibinfo{person}{Linying Jiang}, {and} \bibinfo{person}{Xingwei Wang}.} \bibinfo{year}{2023}\natexlab{b}.
\newblock \showarticletitle{ID Embedding as Subtle Features of Content and Structure for Multimodal Recommendation}.
\newblock \bibinfo{journal}{\emph{ArXiv}} (\bibinfo{year}{2023}), \bibinfo{pages}{1--11}.
\newblock


\bibitem[Liu et~al\mbox{.}(2023a)]%
        {pane-gnn}
\bibfield{author}{\bibinfo{person}{Ziyang Liu}, \bibinfo{person}{Chaokun Wang}, \bibinfo{person}{Jingcao Xu}, \bibinfo{person}{Cheng Wu}, \bibinfo{person}{Kai Zheng}, \bibinfo{person}{Yang Song}, \bibinfo{person}{Na Mou}, {and} \bibinfo{person}{Kun Gai}.} \bibinfo{year}{2023}\natexlab{a}.
\newblock \showarticletitle{PANE-GNN: Unifying Positive and Negative Edges in Graph Neural Networks for Recommendation}.
\newblock \bibinfo{journal}{\emph{arXiv preprint arXiv:2306.04095}} (\bibinfo{year}{2023}).
\newblock


\bibitem[Mao et~al\mbox{.}(2021)]%
        {ultragcn}
\bibfield{author}{\bibinfo{person}{Kelong Mao}, \bibinfo{person}{Jieming Zhu}, \bibinfo{person}{Xi Xiao}, \bibinfo{person}{Biao Lu}, \bibinfo{person}{Zhaowei Wang}, {and} \bibinfo{person}{Xiuqiang He}.} \bibinfo{year}{2021}\natexlab{}.
\newblock \showarticletitle{UltraGCN: Ultra Simplification of Graph Convolutional Networks for Recommendation}. In \bibinfo{booktitle}{\emph{Proceedings of the 30th ACM International Conference on Information \& Knowledge Management}}. \bibinfo{pages}{1253–1262}.
\newblock


\bibitem[Rong et~al\mbox{.}(2020)]%
        {graphtransformer}
\bibfield{author}{\bibinfo{person}{Yu Rong}, \bibinfo{person}{Yatao Bian}, \bibinfo{person}{Tingyang Xu}, \bibinfo{person}{Weiyang Xie}, \bibinfo{person}{Ying WEI}, \bibinfo{person}{Wenbing Huang}, {and} \bibinfo{person}{Junzhou Huang}.} \bibinfo{year}{2020}\natexlab{}.
\newblock \showarticletitle{Self-Supervised Graph Transformer on Large-Scale Molecular Data}. In \bibinfo{booktitle}{\emph{Advances in Neural Information Processing Systems}}. \bibinfo{pages}{12559--12571}.
\newblock


\bibitem[Seo et~al\mbox{.}(2022)]%
        {siren}
\bibfield{author}{\bibinfo{person}{Changwon Seo}, \bibinfo{person}{Kyeong-Joong Jeong}, \bibinfo{person}{Sungsu Lim}, {and} \bibinfo{person}{Won-Yong Shin}.} \bibinfo{year}{2022}\natexlab{}.
\newblock \showarticletitle{SiReN: Sign-Aware Recommendation Using Graph Neural Networks}.
\newblock \bibinfo{journal}{\emph{IEEE Transactions on Neural Networks and Learning Systems}} (\bibinfo{year}{2022}), \bibinfo{pages}{1--15}.
\newblock


\bibitem[Shu et~al\mbox{.}(2021)]%
        {sgcl}
\bibfield{author}{\bibinfo{person}{Lin Shu}, \bibinfo{person}{Erxin Du}, \bibinfo{person}{Yaomin Chang}, \bibinfo{person}{Chuan Chen}, \bibinfo{person}{Zibin Zheng}, \bibinfo{person}{Xingxing Xing}, {and} \bibinfo{person}{Shaofeng Shen}.} \bibinfo{year}{2021}\natexlab{}.
\newblock \showarticletitle{SGCL: Contrastive Representation Learning for Signed Graphs}.
\newblock \bibinfo{journal}{\emph{Proceedings of the 30th ACM International Conference on Information \& Knowledge Management}} (\bibinfo{year}{2021}), \bibinfo{pages}{1671–1680}.
\newblock


\bibitem[Velickovic et~al\mbox{.}(2017)]%
        {gat}
\bibfield{author}{\bibinfo{person}{Petar Velickovic}, \bibinfo{person}{Guillem Cucurull}, \bibinfo{person}{Arantxa Casanova}, \bibinfo{person}{Adriana Romero}, \bibinfo{person}{Pietro Lio’}, {and} \bibinfo{person}{Yoshua Bengio}.} \bibinfo{year}{2017}\natexlab{}.
\newblock \showarticletitle{Graph Attention Networks}.
\newblock \bibinfo{journal}{\emph{ArXiv}} (\bibinfo{year}{2017}), \bibinfo{pages}{1--12}.
\newblock


\bibitem[Wang et~al\mbox{.}(2017)]%
        {sine}
\bibfield{author}{\bibinfo{person}{Suhang Wang}, \bibinfo{person}{Jiliang Tang}, \bibinfo{person}{Charu~C. Aggarwal}, \bibinfo{person}{Yi Chang}, {and} \bibinfo{person}{Huan Liu}.} \bibinfo{year}{2017}\natexlab{}.
\newblock \showarticletitle{Signed Network Embedding in Social Media}. In \bibinfo{booktitle}{\emph{Proceedings of the 2017 SIAM International Conference on Data Mining}}. \bibinfo{pages}{327--335}.
\newblock


\bibitem[Wang et~al\mbox{.}(2019)]%
        {ngcf}
\bibfield{author}{\bibinfo{person}{Xiang Wang}, \bibinfo{person}{Xiangnan He}, \bibinfo{person}{Meng Wang}, \bibinfo{person}{Fuli Feng}, {and} \bibinfo{person}{Tat-Seng Chua}.} \bibinfo{year}{2019}\natexlab{}.
\newblock \showarticletitle{Neural Graph Collaborative Filtering}. In \bibinfo{booktitle}{\emph{Proceedings of the 42nd International ACM SIGIR Conference on Research and Development in Information Retrieval}}. \bibinfo{pages}{165–174}.
\newblock


\bibitem[Wang et~al\mbox{.}(2023)]%
        {Wang2023LearningFN}
\bibfield{author}{\bibinfo{person}{Yueqi Wang}, \bibinfo{person}{Yoni Halpern}, \bibinfo{person}{Shuo Chang}, \bibinfo{person}{Jingchen Feng}, \bibinfo{person}{Elaine~Ya Le}, \bibinfo{person}{Longfei Li}, \bibinfo{person}{Xujian Liang}, \bibinfo{person}{Min-Cheng Huang}, \bibinfo{person}{Shane Li}, \bibinfo{person}{Alex Beutel}, \bibinfo{person}{Yaping Zhang}, {and} \bibinfo{person}{Shuchao Bi}.} \bibinfo{year}{2023}\natexlab{}.
\newblock \showarticletitle{Learning from Negative User Feedback and Measuring Responsiveness for Sequential Recommenders}. In \bibinfo{booktitle}{\emph{Proceedings of the 17th ACM Conference on Recommender Systems}}. \bibinfo{pages}{1049–1053}.
\newblock


\bibitem[wei Wei et~al\mbox{.}(2019)]%
        {mmgcn}
\bibfield{author}{\bibinfo{person}{Yin wei Wei}, \bibinfo{person}{Xiang Wang}, \bibinfo{person}{Liqiang Nie}, \bibinfo{person}{Xiangnan He}, \bibinfo{person}{Richang Hong}, {and} \bibinfo{person}{Tat-Seng Chua}.} \bibinfo{year}{2019}\natexlab{}.
\newblock \showarticletitle{MMGCN: Multi-modal Graph Convolution Network for Personalized Recommendation of Micro-video}. In \bibinfo{booktitle}{\emph{Proceedings of the 27th ACM International Conference on Multimedia}}. \bibinfo{pages}{1437--1445}.
\newblock


\bibitem[Wu et~al\mbox{.}(2019)]%
        {SimplifyingGCN}
\bibfield{author}{\bibinfo{person}{Felix Wu}, \bibinfo{person}{Tianyi Zhang}, \bibinfo{person}{Amauri~H. de Souza}, \bibinfo{person}{Christopher Fifty}, \bibinfo{person}{Tao Yu}, {and} \bibinfo{person}{Kilian~Q. Weinberger}.} \bibinfo{year}{2019}\natexlab{}.
\newblock \showarticletitle{Simplifying Graph Convolutional Networks}. In \bibinfo{booktitle}{\emph{International Conference on Machine Learning}}. \bibinfo{pages}{6861--6871}.
\newblock


\bibitem[Wu et~al\mbox{.}(2021)]%
        {wu2021self}
\bibfield{author}{\bibinfo{person}{Jiancan Wu}, \bibinfo{person}{Xiang Wang}, \bibinfo{person}{Fuli Feng}, \bibinfo{person}{Xiangnan He}, \bibinfo{person}{Liang Chen}, \bibinfo{person}{Jianxun Lian}, {and} \bibinfo{person}{Xing Xie}.} \bibinfo{year}{2021}\natexlab{}.
\newblock \showarticletitle{Self-supervised graph learning for recommendation}. In \bibinfo{booktitle}{\emph{Proceedings of the 44th international ACM SIGIR conference on research and development in information retrieval}}. \bibinfo{pages}{726--735}.
\newblock


\bibitem[Wu et~al\mbox{.}(2022)]%
        {wu2022graph}
\bibfield{author}{\bibinfo{person}{Shiwen Wu}, \bibinfo{person}{Fei Sun}, \bibinfo{person}{Wentao Zhang}, \bibinfo{person}{Xu Xie}, {and} \bibinfo{person}{Bin Cui}.} \bibinfo{year}{2022}\natexlab{}.
\newblock \showarticletitle{Graph neural networks in recommender systems: a survey}.
\newblock \bibinfo{journal}{\emph{Comput. Surveys}} (\bibinfo{year}{2022}), \bibinfo{pages}{1--37}.
\newblock


\bibitem[Xu et~al\mbox{.}(2023)]%
        {Xu2023SignedNetwork}
\bibfield{author}{\bibinfo{person}{Pinghua Xu}, \bibinfo{person}{Wenbin Hu}, \bibinfo{person}{Jia Wu}, \bibinfo{person}{Weiwei Liu}, \bibinfo{person}{Yang Yang}, {and} \bibinfo{person}{Philip~S. Yu}.} \bibinfo{year}{2023}\natexlab{}.
\newblock \showarticletitle{Signed Network Representation by Preserving Multi-Order Signed Proximity}.
\newblock \bibinfo{journal}{\emph{IEEE Transactions on Knowledge and Data Engineering}} (\bibinfo{year}{2023}), \bibinfo{pages}{3087--3100}.
\newblock


\bibitem[Xu et~al\mbox{.}(2022)]%
        {Xu2022Dualbranch}
\bibfield{author}{\bibinfo{person}{Pinghua Xu}, \bibinfo{person}{Yibing Zhan}, \bibinfo{person}{Liu Liu}, \bibinfo{person}{Baosheng Yu}, \bibinfo{person}{Bo Du}, \bibinfo{person}{Jia Wu}, {and} \bibinfo{person}{Wenbin Hu}.} \bibinfo{year}{2022}\natexlab{}.
\newblock \showarticletitle{Dual-branch Density Ratio Estimation for Signed Network Embedding}. In \bibinfo{booktitle}{\emph{Proceedings of the ACM Web Conference 2022}}. \bibinfo{pages}{1651–1662}.
\newblock


\bibitem[Yu et~al\mbox{.}(2020)]%
        {yu2020enhancing}
\bibfield{author}{\bibinfo{person}{Junliang Yu}, \bibinfo{person}{Hongzhi Yin}, \bibinfo{person}{Jundong Li}, \bibinfo{person}{Min Gao}, \bibinfo{person}{Zi-Liang Huang}, {and} \bibinfo{person}{Li zhen Cui}.} \bibinfo{year}{2020}\natexlab{}.
\newblock \showarticletitle{Enhancing Social Recommendation With Adversarial Graph Convolutional Networks}.
\newblock \bibinfo{journal}{\emph{IEEE Transactions on Knowledge and Data Engineering}} (\bibinfo{year}{2020}), \bibinfo{pages}{3727--3739}.
\newblock


\bibitem[Zhang et~al\mbox{.}(2021)]%
        {lattice}
\bibfield{author}{\bibinfo{person}{Jinghao Zhang}, \bibinfo{person}{Yanqiao Zhu}, \bibinfo{person}{Qiang Liu}, \bibinfo{person}{Shu Wu}, \bibinfo{person}{Shuhui Wang}, {and} \bibinfo{person}{Liang Wang}.} \bibinfo{year}{2021}\natexlab{}.
\newblock \showarticletitle{Mining Latent Structures for Multimedia Recommendation}. In \bibinfo{booktitle}{\emph{Proceedings of the 29th ACM International Conference on Multimedia}}. \bibinfo{pages}{3872–3880}.
\newblock


\bibitem[Zhang et~al\mbox{.}(2023)]%
        {sbgcl}
\bibfield{author}{\bibinfo{person}{Zeyu Zhang}, \bibinfo{person}{Jiamou Liu}, \bibinfo{person}{Kaiqi Zhao}, \bibinfo{person}{Song Yang}, \bibinfo{person}{Xianda Zheng}, {and} \bibinfo{person}{Yifei Wang}.} \bibinfo{year}{2023}\natexlab{}.
\newblock \showarticletitle{Contrastive Learning for Signed Bipartite Graphs}. In \bibinfo{booktitle}{\emph{Proceedings of the 46th International ACM SIGIR Conference on Research and Development in Information Retrieval}}. \bibinfo{pages}{1629–1638}.
\newblock


\end{thebibliography}

\appendix

\end{document}